\documentclass[journal,10pt]{IEEEtran}
\usepackage[subpreambles=false]{standalone} %enable standalone subfiles should be placed early in the document.
\usepackage{import}
\usepackage{graphicx}
\usepackage{subfigure}
\usepackage{float} %tell latex I REALLY want the figure in that specific place, no matter how much whitespace will be left
\usepackage{textcomp} %supports the Text Companion fonts, which provide many text symbols
\usepackage{cite} %Multiple citations in single cite command
\usepackage{amsmath,amssymb,amsfonts} %IEEE
\usepackage{amsthm} %facilitates the kind of theorem setup
\usepackage{mathrsfs} %support some fonts in mathematics
\usepackage[mathscr]{euscript} %provide spacial capital letters
\usepackage{epstopdf} 
\usepackage{xcolor}
\usepackage{pgfplots}
\pgfplotsset{compat=newest} %for double ylabels on the right and the left
\usepackage{tikz} %for best figures ever
\usetikzlibrary{plotmarks}
\usepackage{gensymb} %Provides generic commands \degree, \celsius, \perthousand, \micro
\usetikzlibrary{intersections,backgrounds, patterns}
\usepackage{array}
\usepackage[nolist]{acronym} %for abbreviations
\usepackage{multirow} % for tables
\usepackage{enumerate}
\usepackage{balance}
\usepackage{booktabs}
\usepackage[colorlinks,citecolor=violet,linkcolor=violet,urlcolor=black]{hyperref}%使用该宏包后会自动生成书签，引用位置会有颜色标出。

\begin{document}

\title{Edge Computing Assisted Autonomous Flight for UAV: Synergies between Vision and Communications}

\author{Quan~Chen\IEEEauthorrefmark{1}\IEEEauthorrefmark{2},~
Hai~Zhu\IEEEauthorrefmark{3},~
Lei~Yang\IEEEauthorrefmark{1},
Xiaoqian~Chen\IEEEauthorrefmark{1},
Sofie~Pollin\IEEEauthorrefmark{2},
and~Evgenii~Vinogradov\IEEEauthorrefmark{2}%
\\

\IEEEauthorrefmark{1}College of Aerospace Science and Engineering, National University of Defense Technology\\
\IEEEauthorrefmark{2}Department of Electrical Engineering, KU Leuven\\
\IEEEauthorrefmark{3}Department of Cognitive Robotics, Delft University of Technology\\
}
%\markboth{IEEE Communications Magazine,~Vol.~0, No.~0, 2020}%
\maketitle

\begin{abstract}

Autonomous flight for UAVs relies on visual information for avoiding obstacles and ensuring a safe collision-free flight. In addition to visual clues, safe UAVs often need connectivity with the ground station. In this paper, we study the synergies between vision and communications for edge computing-enabled UAV flight. 
By proposing a framework of Edge Computing Assisted Autonomous Flight (ECAAF), we illustrate that vision and communications can interact with and assist each other with the aid of edge computing and offloading, and further speed up the UAV mission completion. ECAAF consists of three functionalities that are discussed in detail: edge computing for 3D map acquisition, radio map construction from the 3D map, and online trajectory planning. During ECAAF, the interactions of communication capacity, video offloading, 3D map quality, and channel state of the trajectory form a positive feedback loop. Simulation results verify that the proposed method can improve mission performance by enhancing connectivity. Finally, we conclude with some future research directions. 
\end{abstract}

\IEEEpeerreviewmaketitle

\section{Introduction\label{section:introduction}}
With the advantages of high mobility, flexible deployment, and low cost, unmanned aerial vehicles (UAVs), or drones, have drawn great attention in commercial, civil, and military fields. Generally, UAVs are expected to quickly complete their mission while guaranteeing autonomy and safety.
Autonomy and safety require good environmental awareness on the one hand, but also reliable wireless connectivity for remote control, cooperation, and aerial deconfliction on the other hand \cite{Vinogradov2019WirelessCF}. As both visual mapping and wireless connectivity are a given, the main question we answer is how they could strengthen each other synergistically with the aid of edge computing.

Current UAV-ground communications mostly adopt short-range local network techniques such as  WiFi \cite{Zeng2019MWC}. Although easy to use, they have some limitations in operating range and co-channel interference management \cite{Challita2019}. Recently, with the rapid development of cellular networks (LTE and 5G), cellular-connected UAV has become a promising solution where the UAV connects with cellular base stations (BSs) as an aerial user along with ground users \cite{Zeng2019}. Cellular networks allow the UAV to maintain connectivity in a large range. The 5G Ultra-Reliable Low-Latency Communication (URLLC) service also meets the demands of low latency, real-time communication, and robust security. 

In some visual UAV applications (e.g., surveillance, monitoring), the onboard camera is the primary payload to perform the task. It also plays a key role in supporting subsystems and is widely adopted by UAVs for autonomous navigation and obstacle avoidance. 
From the captured video, it is possible to extract auxiliary information such as the UAV position and 3D map features of the explored areas \cite{Qin2019}. With this information, the UAV is able to obtain awareness of the environment around it and autonomously adjust its trajectory. Besides, the UAV movement is also affected by the information update rate: a higher flying speed requires a faster update rate, which poses a challenge for the onboard processing capability. However, in the upcoming 5G era, Mobile Edge Computing (MEC) enables powerful computing capability at the edge of cellular networks \cite{Zeng2019}. Similar to autonomous driving, the onboard computational tasks can be offloaded to the ground and processed remotely, which speeds up the information acquisition for UAV \cite{Dey2016}.

On the other hand, the explored environmental information can be exploited to improve communication performance. Specifically, the 3D map enables the UAV to be aware of the obstacles between the BS and UAV that influence the signal propagation. The 3D map can be generated from the captured video by vision-processing algorithms such as simultaneous localization and mapping (SLAM) \cite{Qin2019}. The key idea in this paper is to leverage the captured video to incrementally construct a 3D map of the environment for the UAV and predict the wireless link states around it; then the UAV adaptively designs its real-time trajectory,  trying to maintain Line-of-Sight (LoS) link with the BS. The UAV can offload its computing tasks to the edge computing server for faster video processing and map acquisition. Since typically the speed of a vision-navigated UAV is limited by the map update rate, edge computing also enhances UAV mobility and makes the trajectory more flexible. Moreover, edge computing relies on the offloading link capacity. Thus, in turn, a good link guarantees high-quality mapping and UAV mobility. By the positive interaction of visual information and communications, both of them can be enhanced. 

Inspired by the above idea, in this paper, we propose a framework of Edge Computing Assisted Autonomous Flight (ECAAF) for UAV flying in urban areas. The contributions can be summarized as follows: i) Focusing on the synergies between vision and communications, we propose a solution to realize real-time path planning for the UAV in urban areas without known maps. ii) By ECAAF, we illustrate that vision and communications can interact with and assist each other with the aid of edge computing and task offloading. The framework architecture and trade-offs are thoroughly discussed. iii) We offer a detailed implementation for ECAAF solution and verify it by a case study.

\section{Communications and vision for UAV flight}\label{section:SoA}
This section presents the communication and vision related approaches to improve UAV flight performance. We discuss two specific techniques that are typically considered in isolation: communication-aware, and vision-assisted trajectory planning. For autonomous flight, communication awareness requires the UAV to sense or predict its link states and keep the connectivity reliable. And vision awareness obtained from SLAM enables the UAV to construct and update the 3D map of its environment using its onboard sensors, so that it can adaptively adjust its trajectory.

\subsection{Communication-aware flight\label{section:SoA1}}
In order to maintain a good connection with the BS, the UAV trajectory or placement can be deliberately designed with communication awareness. There are several approaches to enhance communication awareness, such as LoS-based channel modeling, channel measurement-based estimation, and map-assisted channel estimation \cite{Zeng2019}.

For the channel modeling, unexpected obstacles are the primary cause of the degradation and uncertainty of air-to-ground link, especially in urban areas. Due to the obstacles, the link state can be either LoS or NLoS (Non-LoS). The latter results in much lower received signal quality. To estimate the channel gain, a widely adopted method is to approximate link path loss (i.e., signal attenuation in the channel) by estimating the LoS occurrence probability which depends on the UAV-ground elevation angle \cite{You2019}. Some studies adopt this model to define the UAV altitude for maximal coverage \cite{Al-Hourani2016}, and to optimize the UAV trajectory for maximal data collection in a UAV-enabled wireless sensor network (WSN) application \cite{You2019}. 
The statistical channel models are general and provide good insights to the network designers. However, the LoS-probability-based method does not apply to real-time flights in obstacle-dense regions. One improved method is to combine the offline LoS modeling and online channel measurement-based schemes \cite{You2020}: firstly, an offline path is optimized using the probabilistic LoS model; next, the UAV adaptively adjusts its speed and communication scheduling along the offline path based on the instantaneous channel state information (CSI). Although the leaning-based CSI approach can infer the LoS state from the variation pattern of CSI over time, this approach is mainly effective at the measured moments and locations, which limits its application in large and unexplored regions. If learning approach is further adopted to predict the CSI map of unexplored regions, it requires massive offline training data and dedicated training model \cite{You2020}, which might be unavailable in a new environment.

Another way to estimate the channel gain is by exploiting the real 3D map instead of the above two approaches. The map-based LoS/NLoS classification works for larger regions than the CSI approach, and it is more reliable for locations where the UAV has never been. Once the 3D map is given, the link state (LoS or NLoS) between any two locations can be accurately inferred \cite{Zeng2019}. This map-aided method has been implemented in several UAV applications. For example, in \cite{Esrafilian2020}, by combining the 3D map and a path loss model to generate the coverage map, the shortest trajectory is designed under cellular connectivity constraints. Note that the premise of this method is to obtain a priori 3D map of the flight-related regions, which may be not available or reliable in practice.

\subsection{Vision-assisted flight\label{section:SoA2}}
Vision is another type of important supporting information for UAV flight that can be captured by onboard vision-sensors (e.g., depth camera). 
To extract valid information from the captured video, SLAM is a widely adopted technique where the UAV is capable of online map building, while simultaneously utilizing the generated map to estimate and correct errors in the navigation solution obtained \cite{Kim2007}. When video frames are captured, they are first pre-processed, e.g., pose feature extraction, pose estimation, and keyframe selection. Then based on the keyframes, visual tracking and dense map fusion are executed. Finally, the updated 3D map can be obtained after a loop correction \cite{Zheng2020}.
With the map information, the UAV can not only recognize the obstacles but also plan its trajectory to avoid them.
Therefore, SLAM enables the UAV to autonomously and safely fly in map-unavailable areas and even GPS-denied areas \cite{Qin2019}.

However, real-time SLAM requires a huge amount of computations, especially when the vehicle is moving at high speed which requires a high SLAM update rate. This requirement quickly exceeds the limited onboard computing power of UAV and limits its flying speed and flexibility. A potential solution is to offload the computational task to the ground server and send the obtained SLAM results back to the UAV. This mode is known as cloud SLAM that has been studied in some practical applications \cite{Zheng2020}. Robo-Earth, as a representative project, has successfully demonstrated some cloud SLAM cases \cite{Riazuelo2015} in practice. Note that here the 'cloud' concept only refers to non-local processing, which is not identical to that in 'cloud computing.' In cellular networks, cloud SLAM requires extra computing capability at BS side, which just matches the typical scenario of the upcoming 5G with MEC \cite{Zeng2019}.

\begin{figure*}[ht!]
\centering
\includegraphics[width=0.7\linewidth]{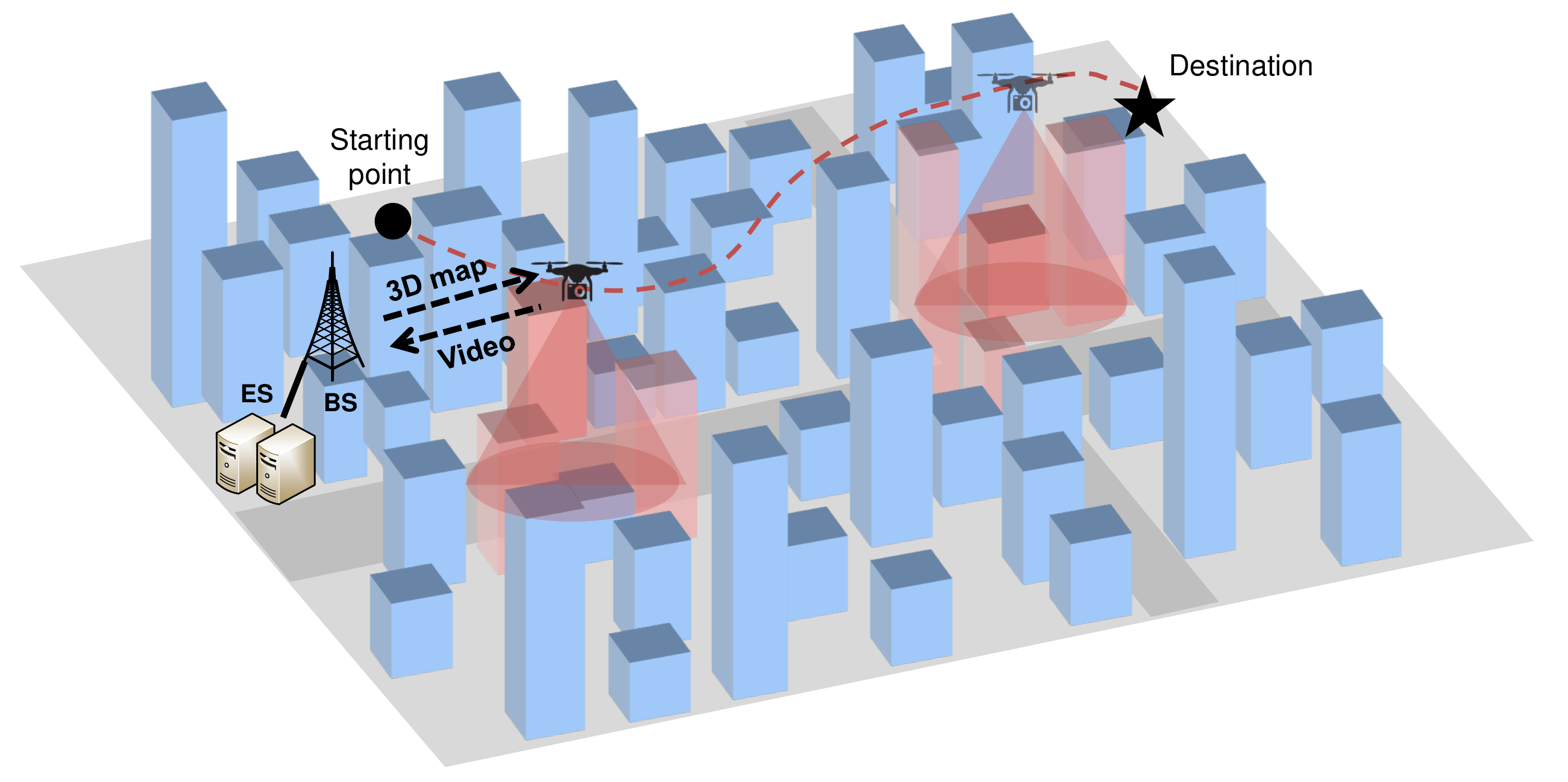}
\caption{Edge computing assisted autonomous flight (ECAAF). The UAV offloads its captured video to the Edge Server and receives the processed 3D map. The awareness of the local environment helps the UAV maintain a LoS link with BS by online trajectory planning.}
\label{fig:3D city UAV model}
\end{figure*}

\section{ECAAF framework}\label{section:ECAAF}
According to the above introduction, both awareness of communication and vision can promote UAV mission performance. In this section, by combining the two approaches and enabling SLAM offloading, we propose a framework of edge computing assisted autonomous flight (ECAAF) to interactively improve SLAM and communication performances. 

\subsection{Features of ECAAF: vision and communication synergies}\label{section:ECAAF features}
\begin{figure}[ht!]
\centering
\includegraphics[width=0.9\linewidth]{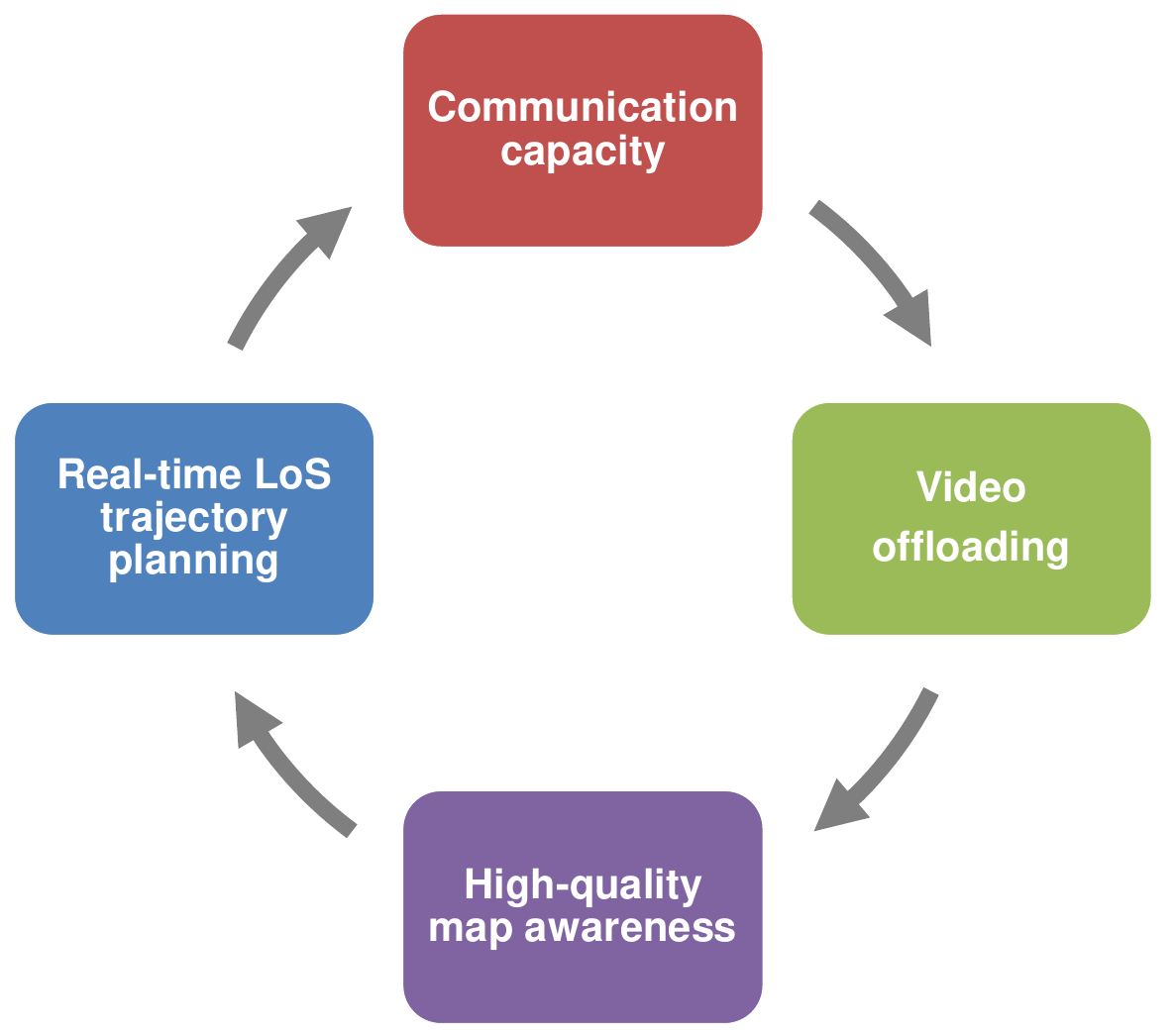}
\caption{Positive interaction loop of vision and communications in ECAAF. Better communication improves environment perception, which in turn boosts communication performance.}
\label{fig:Interactive loop}
\end{figure}
As illustrated in Fig. \ref{fig:3D city UAV model}, a cellular-connected UAV keeps communication with a cellular BS during the whole flight. The UAV captures real-time video and can offload it to the BS. The BS is connected with an Edge Server (ES) where the offloaded SLAM task is processed. Next, the obtained SLAM results are sent back to the UAV for map construction and online trajectory planning. The goal is to design a real-time trajectory from Starting point to Destination with i) minimum duration; and ii) less time and distance where the UAV-BS link is NLoS (i.e., unreliable channel).

During the ECAAF, visual perception and communications interact and promote each other, as shown in Fig. \ref{fig:Interactive loop}. A higher-capacity link allows high-speed video transmission between UAV and BS, which improves the 3D map quality and update rate. Then the UAV obtains a better awareness of its wireless environment. It can also fly faster with a higher update rate of mapping and localization. Accordingly, the UAV can more accurately plan its real-time trajectory with better LoS states, which in turn promotes the communication capacity and reliability. Therefore, a positive interactive loop is formed where better communication improves environment perception, which meanwhile boosts communication performance.

\subsection{Architecture of ECAAF}\label{section:ECAAF architecture}

\begin{figure*}[ht!]
\centering
\includegraphics[width=0.7\linewidth]{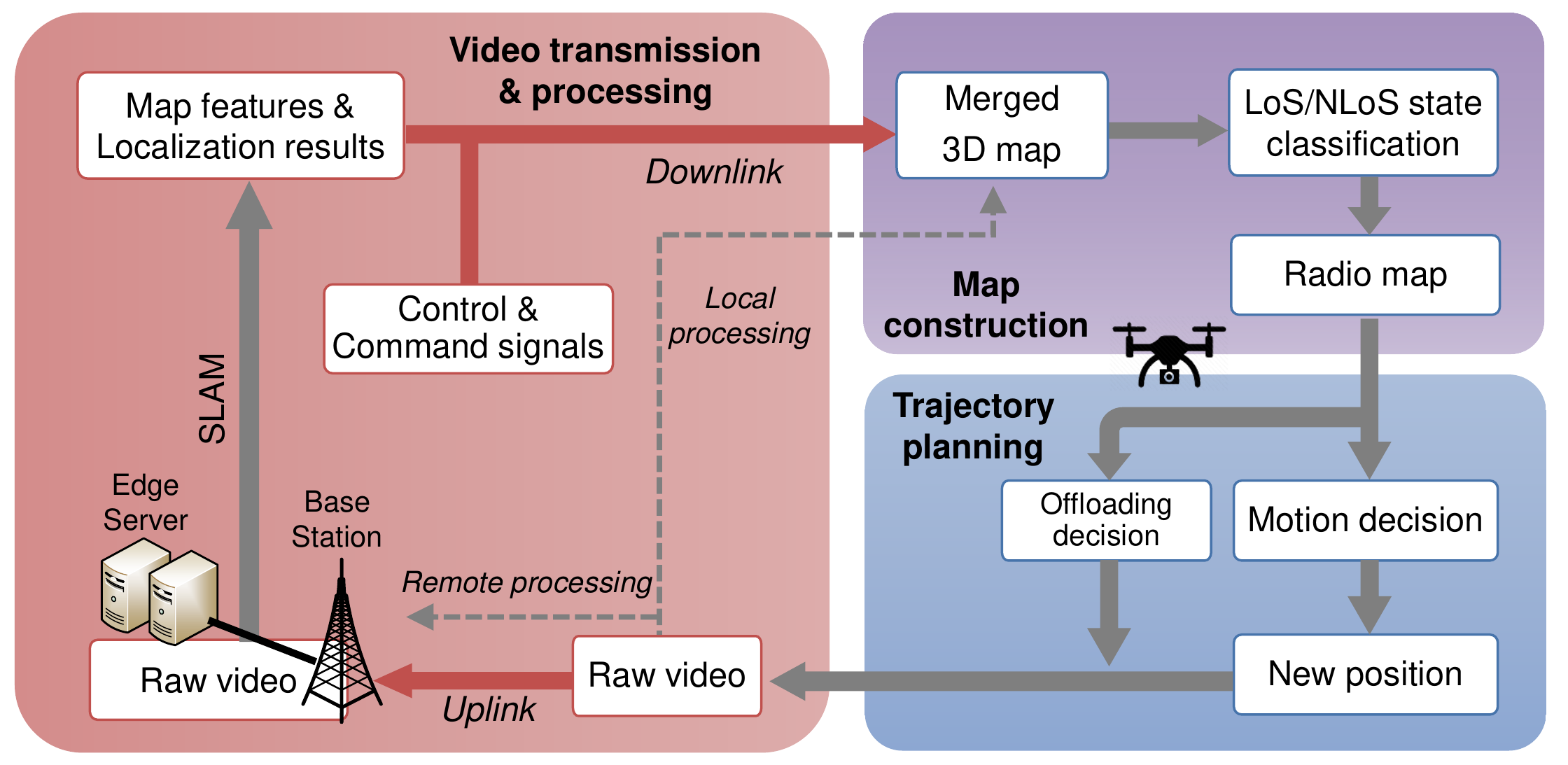}
\caption{ECAAF architecture. The captured video is offloaded to cellular BS and processed at ES. Then the obtained SLAM results are sent back to the UAV and help it perceive its wireless environment and plan its real-time trajectory. The data flow forms a closed loop.}
\label{fig:ECAAF architecture}
\end{figure*}

Fig. \ref{fig:ECAAF architecture} shows the architecture of ECAAF, consisting of three parts:

i) Video transmission and processing: When the raw video is captured, it is compressed and offloaded to the UAV’s serving BS through uplink. Then, the map features are extracted at the ES side by the SLAM algorithm. 

ii) Map construction: The obtained map features and localization results are sent back to the UAV via downlink. Then the UAV merges new map features into the existing local 3D map and transforms it into a radio map (a distribution of uplink channel gain) by classifying the link state of neighboring locations. 

iii) Trajectory planning: From the map, the UAV knows which directions are promising to achieve higher link capacity. It accordingly plans its following trajectory as well as the offloading decisions (i.e., whether the video is processed locally or remotely). When UAV moves to a new position, the incoming video is processed using the same process.

From the communication aspect, the 3D map acquisition relies on both downlink and uplink. The raw video offloading is through uplink and the generated map features transmission is through downlink. The uplink determines the offloading rate and thus affects the remote processing rate, while the requirements for downlink are lower because the returned map features are much less data.
The downlink also transmits command and control signals from the ground mission control station for mission control and necessary intervention.

The ECAAF implementation has the following requirements which have been separately realized in practical cases \cite{Zeng2019,Kim2007, Riazuelo2015, Zheng2020}: i)  cellular network connectivity; ii) an onboard visual sensor (monocular or stereo camera) to capture video; iii) an ES and onboard computer with SLAM solving capability.

\section{Implementation of ECAAF}\label{section:ECAAF implementation}
This section concentrates on the specific technical details of ECAAF solution: radio map construction, flying speed adjustment, and online trajectory planning. The details of SLAM algorithms can be found in \cite{Kim2007} and are not described here. 
\subsection{Radio map construction}\label{section:radio map construction}
Upon receiving the new 3D map features from the ES, the UAV merges them into the local map and updates its stored map, which contains the building height data of the explored areas. Then the UAV classifies the link state of its neighboring locations and then transforms the obstacle map into a radio map.

\begin{figure*}[t!]
\centering 
\includegraphics[width=0.7\linewidth]{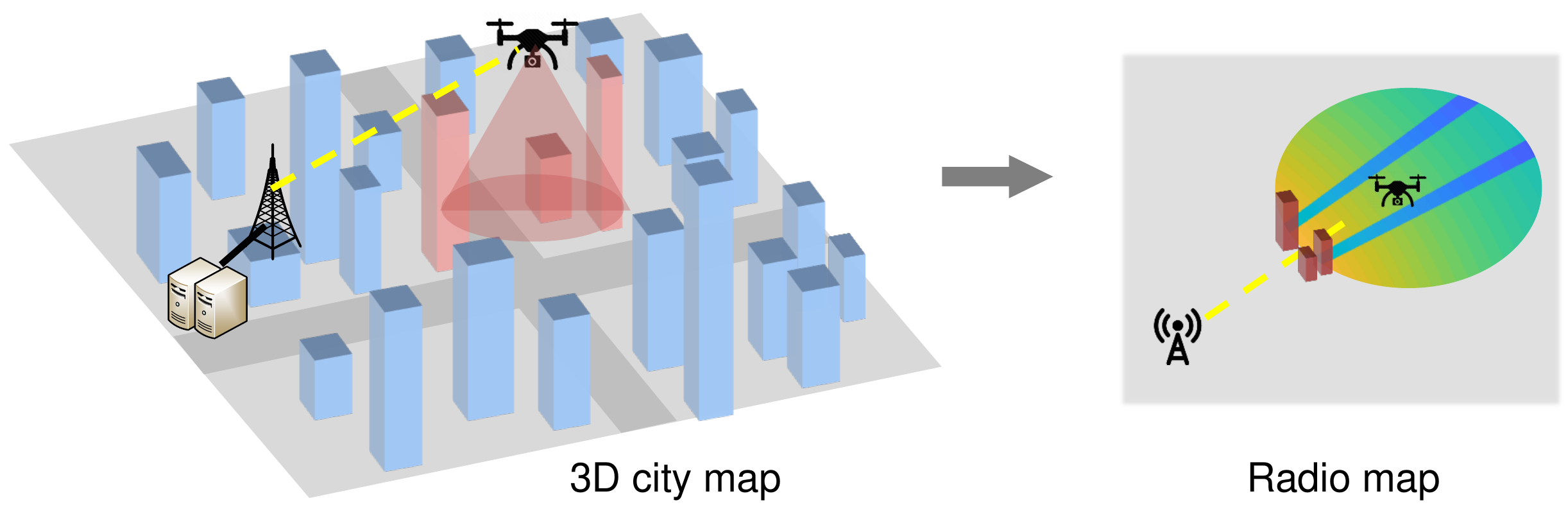} 
\caption{Map transformation from the 3D map to a radio map. The radio map refers to the distribution of channel gain, which is affected by the obstacles.}
\label{fig:Map transformation}
\end{figure*}

\textbf{Link state classification:} 
Knowing the BS location and UAV current location, the UAV-BS direct line can be calculated. The UAV judges whether the line intersects with obstacles in its explored map and classifies the link state as LoS (no intersection) or NLoS (intersection occurs), which is marked by a state indicator.

On the other hand, the BS can infer the LoS/NLoS states of the UAV position by analyzing the variation pattern of the measured CSI. Then the obtained LoS/NLoS information is sent to the UAV and helps the UAV to correct the map-based link state classification. Because the complete map is unknown to the UAV, there may be unexplored regions lacking map information along the UAV-BS line. In this case, those map-missing parts are first assumed to be LoS and then corrected by the CSI-based classification. If the link states of some directions have been marked as NLoS earlier, they keep the NLoS state without further estimation.

\textbf{Map transformation:}
After obtaining the link state of current location and surroundings, the UAV calculates the path loss and channel gain at each estimated location and saves it to a radio map. The radio map is represented by a 3-dimensional table which saves the channel gain corresponding to each location. The locations are defined in a BS-centered coordinate system. When UAV explores new space, the channel gain is calculated and added to the table. In this way, the 3D vision map is transformed into a 3D radio map. Fig.~\ref{fig:Map transformation} gives a diagram of the map transformation where a horizontal 2D slice of radio map is given for simplicity.

\subsection{Flying speed adjustment}\label{section:flying speed adjustment}
Because of the requirements of control accuracy and flight safety, UAV speed should be kept within a safe range, which depends on the offloading link capacity. In a simplified model, let the maximum speed be proportional to the mapping \& localization update rate,
which can be viewed as the frame rate of SLAM processing and is limited by offloading transmission (remote processing time is ignored). The achievable frame rate is determined by link capacity and average data size per frame. Thus, the speed limit also depends on these two values.
For example, assume the required SLAM information for safe movement is 2 frame/m and the file size of each frame is 1~Mbit. When the offloading link capacity is 10 Mbps; If the transmission delay is solely considered, the achievable frame update rate is 10 frame/s, and the UAV speed can achieve a maximum 5 m/s. Thus, the UAV should adaptively adjust its speed of the following trajectory based on the link capacity and safe flight requirements.
Note that the remote processing mode also causes additional delay for the results feedback loop and local action decision, which will lower the map update rate. The additional delay includes processing delay and uplink/downlink data transmission delay. If certain downlink message is missing, it also lowers the map update rate as well as the speed limitation. When the link state is bad and UAV switches to local processing mode, the speed limitation depends on the local computing performance. 

\subsection{Online trajectory planning}\label{section:online trajectory planning}
At each moment, with the incrementally constructed environment map and the corresponding transformed radio map, a local LoS-aware and collision-free trajectory is planned online for the UAV using model predictive control (MPC) \cite{camacho2013model}. MPC has been widely used for online trajectory generation of aerial vehicles \cite{zhu2020icra}. It formulates a constrained optimization problem with a local short planning horizon, in which the LoS-dependent speed constraints and the environment-dependent obstacle avoidance constraints are imposed. The optimal local trajectory is then obtained by minimizing the flying time of the trajectory while satisfying the imposed constraints. The above formulated constrained optimization problem is solved at each moment in real-time, with updated environment map and radio map, until the UAV reaches its destination.

\section{Case study}\label{section:case study}
In this section, by comparison, we give a representative case to illustrate the performance improvement of the proposed ECAAF approach, then present the statistical results of multiple tests.

\subsection{Scenario settings}\label{section:scenario settings}

\begin{figure*}[ht]
    \centering
    \includegraphics[width=0.7\linewidth]{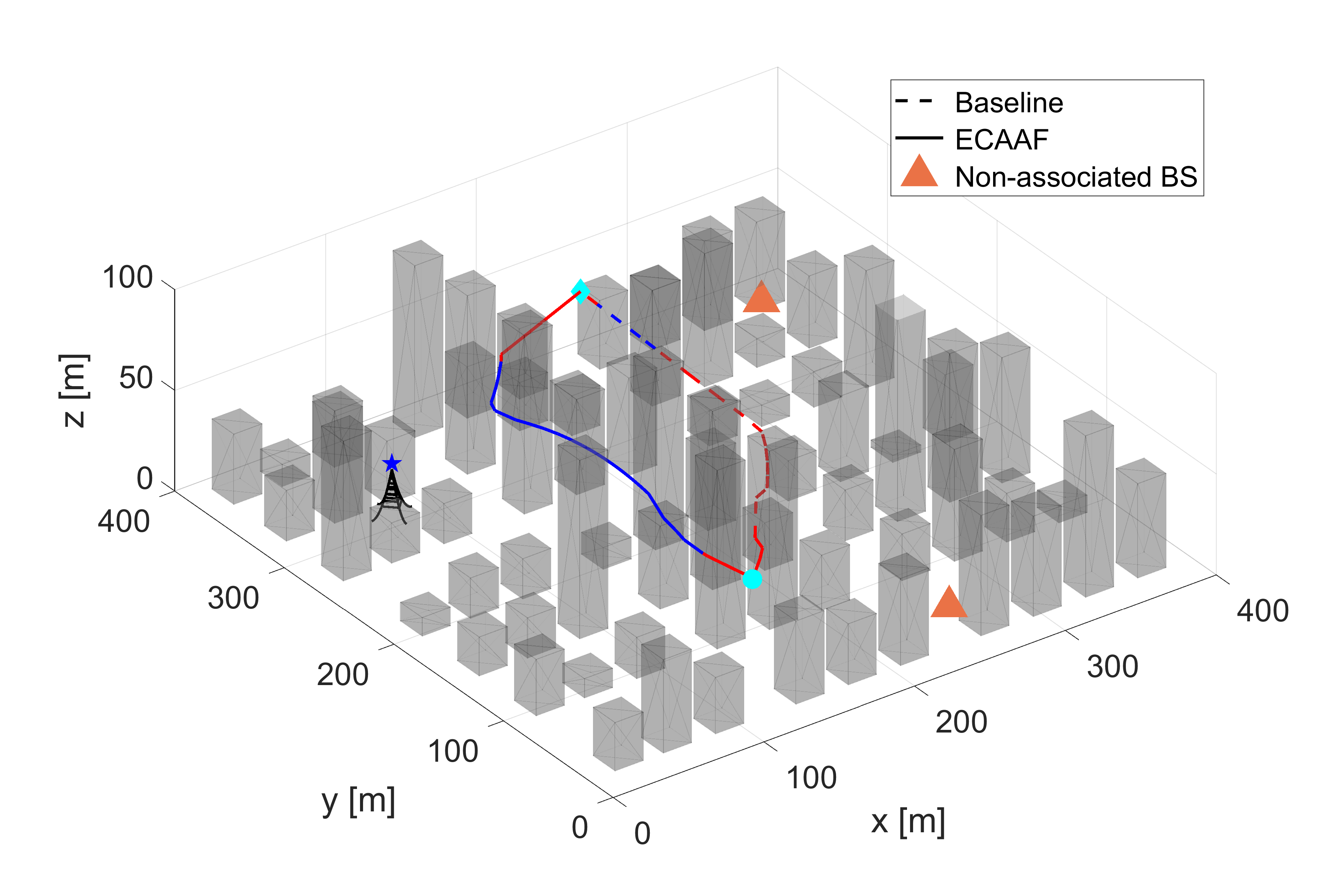}
    \caption{Trajectories of ECAAF (left solid line) and baseline without LoS awareness (right dashed line). Blue and red color indicate LoS and NLoS parts, respectively. The BS with blue star represents the associated BS. Gray boxes are buildings. }%
    \label{fig:trajectory}%
\end{figure*}

We consider a 400\( \times \)400$\text{m}^2$ city map with dense Manhattan-like building distribution. The building height follows the Rayleigh distribution \cite{Al-Hourani2016} with a scale parameter of 35~m. 
The distance between the start and end points is 320m. 
The maximum flying speed and altitude of UAV are 15 m/s and 50m, respectively. Three BSs are deployed in the scenario and the UAV maintains connection with one BS as shown in Fig.~\ref{fig:trajectory}.

The BS height is 25m. 
The UAV antenna adopts a directional pattern as in \cite{Zeng2019} and the transmit power is 30 dBm. Other detailed settings and parameters can be found in the simulation setup in \cite{Zeng2019MWC}. The required SLAM update rate for safe movement is 2 frame/m. The offloading model and SLAM processing settings refer to \cite{Dey2016}, and the camera field of view is set as 120 deg. In the online trajectory planning phase, the planning horizon is set as 10~seconds, and the algorithm parameters refer to \cite{zhu2020icra}. The objective is to minimize the flight duration as well as NLoS duration with the constraints stated in Section \ref{section:ECAAF}.

For comparison, a baseline trajectory without LoS awareness and an optimal trajectory (denoted as GlobeMap) with global environmental awareness are solved. In the baseline, the link capacity is estimated using the probabilistic LoS model \cite{Al-Hourani2016}, and an extra obstacle avoidance mechanism \cite{zhu2020icra} is added to the trajectory. The baseline only uses the local map to avoid obstacles but not estimates LoS states. In the GlobeMap case, the trajectory is obtained under a known global map for both LoS awareness and obstacle avoidance. The baseline, ECAAF, and GlobeMap correspond to the without map, with self-explored local map, and with global map cases, respectively.

\begin{table*}[ht]
\centering
\caption{Comparison results of ECAAF, baseline and GlobeMap in 20 tests.}
\label{tab:results}
\begin{tabular}{|c|c|c|c|c|}
\hline
         & Total flight distance & Total flight duration & Average uplink capacity & NLoS distance ratio\\ \hline
Baseline & 6.1 km            & 971 s           & 14.8 Mbps                   & 49.2 \%    \\ \hline
ECAAF    & 6.5 km            & 755 s           & 16.3 Mbps                   & 27.3 \%  \\ \hline
GlobeMap    & 6.4 km            & 742 s           & 16.4 Mbps                   & 26.8 \%  \\ \hline
\end{tabular}
\end{table*}

\subsection{Results}\label{section:results}

Fig.~\ref{fig:trajectory} illustrates an ECAAF trajectory compared with the baseline where the LoS and NLoS parts are marked on the trajectories. The trajectory of GlobeMap is much close to ECAAF trajectory and thus not drawn. 
Compared to the baseline, with the environmental awareness, the ECAAF trajectory avoids many obstacles between the UAV and BS. Thus a higher link capacity and faster speed are achieved. The flight distance of the baseline, ECAAF, and GlobeMap is 373m, 402m, and 397m, while the flight duration is respectively 74.5s, 48.3s, and 46.8s. Besides, the average uplink capacity of the three planned trajectories is respectively 13.6 Mbps, 16.7 Mbps, and 16.8 Mbps. In this example, the performance of the proposed ECAAF is much promoted compared to the conventional non-map-aware method, and close to the global-map-aware method. With the local map and LoS awareness, the UAV manages to fly in good-channel regions and achieves higher link capacity. Although the detour causes some extra journeys, the UAV can fly faster with the improved link capacity, thus the total flight duration is significantly reduced. Because the locations of non-associated BSs are not known to the UAV, the UAV does not proactively predict the interference. When it flies towards the non-associated BS, the degraded downlink leads to extra delays and a lower flying speed.

We also further repeat the simulation for 20 times with different start and end points and with the straight-line distance ranging from 200 to 400 m. Before each test, a new map is initialized. Table \ref{tab:results} lists the comparison results of the average performance metrics of the 20 runs between the three types of trajectory. In the table, the NLoS distance ratio is the percentage of NLoS path length in the total flight path. When the UAV falls into NLoS link, the interference shows more significant impacts on the transmission delay and the UAV tends to switch to local processing mode. It can be seen from the table that our proposed ECAAF can achieve faster flying and higher overall uplink capacity while keeping a lower NLoS distance radio compared with the baseline where LoS information is not used in trajectory planning. Besides, although the environmental awareness is local, the performance of ECAAF is close to the GlobeMap approach. 
In some cases, when both ECAAF and GlobeMap maintain LoS with the associated BS, the trajectory in GlobeMap deflects away from the non-associated BSs to mitigate the downlink interference.
Note that in the randomly generated scenarios, sometimes the three approaches can obtain all-LoS trajectories, which causes little difference between their performances and lowers the averaged promotion. 

\section{Potential trade-offs and Future Work}\label{section:future}
The performance of ECAAF framework will be affected by many factors in practice. In this section, we list some potential trade-offs that can be explored in order to achieve even better performance than the current version of the ECAAF framework. We also present future research directions.

\subsection{Potential trade-offs}\label{section:potential trade-offs}

\textbf{Local processing vs. cloud processing:}
As an autonomous system, the UAV also has local computation capability \cite{Kim2007}. Since a higher map update rate relaxes the UAV velocity limitation, in the proposed framework, the UAV always selects the processing mode with the higher achievable map update rate. If UAV falls into a bad communication state, the offloading-constrained cloud processing may be even worse than local processing. Then the UAV stops offloading and switches to local processing mode.
Compared with cloud SLAM, the local SLAM algorithm is simpler and has inferior map quality, but it avoids the delay overhead of uplink/downlink transmission.

\textbf{UAV altitude and map accuracy:}
UAVs may tend to fly higher for better LoS probability, but their altitude is constrained by path loss, regulations, as well as the map accuracy. A higher-altitude means a lower resolution of the explored map, and thus the radio map estimation accuracy and trajectory planning are also affected.

\textbf{Multi-BS interference mitigation:}
In the multi-BS scenario, interference from non-associated BSs also affects the UAV downlink channel quality. The UAV can change its trajectory to avoid the non-associated BSs. But in some cases, the UAV needs to make a trade-off between NLoS link and link with strong interference. On the other hand, in the uplink communication the UAV could also cause interference to those non-associated BSs. If the uplink capacity is sufficient, the
 UAV can flexibly adjust transmit power to mitigate the interference. Moreover, UAVs can also exploit the obstacles as barriers to mitigate interference by dedicated trajectory design. When interference exists, the environment and ECAAF bring new challenges and opportunities for UAV communications.

\textbf{Energy consumption:} When LoS links are achieved, a trajectory with detour may consume extra flying energy. But the good channel also improves the UAV motion flexibility, so that the UAV can fly at an optimal energy-efficient speed for energy saving \cite{Zeng2019}. In an energy-constrained scenario, the path and velocity should be jointly designed.

\subsection{Future research directions}\label{section:future work}

\textbf{Multi-BS handover and cooperation:}
In a multi-BS scenario, the UAV can switch its associated BS for optimal communication during its flight. Thus when designing the trajectory, the UAV should also schedule the handover between different BSs to achieve better LoS channel and avoid interference. On the other hand, the UAV can also benefit from the multi-BS cooperation. The communication resources are jointly optimized across different BSs to avoid co-channel interference and improve link reliability. The multi-BS handover and cooperation lead to new issues such as BS selection, handover scheme, map sharing and migration, and trade-offs between performance and backhaul overhead.

\textbf{Map reuse:}
In this work, we consider the UAV always flying over an unknown environment and thereby always relying on video processing. In some complex missions, the UAV is likely to fly by some explored areas. Then the obtained map can be reused so that the video processing is avoided and the UAV mobility is not limited by the uplink. In these cases, the trajectory and offloading decision can be jointly planned.

\section{Conclusion}\label{section:conclusion}
In this article, a framework of edge computing assisted autonomous flight is proposed to achieve high-capacity and reliable UAV-ground communications. With an onboard camera and cloud SLAM, the UAV is able to explore and exploit the environment and adaptively plan its real-time trajectory for better channel states, which improves the link reliability. The interaction of vision and communications boosts both of them. Finally, the case study verifies the proposed approach. ECAAF is a complex mission in practice. A number of practical factors and trade-offs are worthy of further investigation.

\section{Acknowledgments}
This research is supported by the Research Foundation Flanders (FWO), project no. S003817N (OmniDrone) and the National Natural Science Foundation of China no. 11725211. Q. Chen appreciates the financial aid from China Scholarship Council. Q. Chen and H. Zhu contributed equally to this work.

\balance
\bibliographystyle{IEEEtran}
\bibliography{references}

\section*{Biography}
\footnotesize
\setlength{\parskip}{1em}
\textbf{Quan Chen}
received the B.E. (2015) and M.E. (2017) degrees from National University of Defense Technology (NUDT), China, where he is currently pursuing the PhD degree. He is also a visiting scholar with the Department of Electrical Engineering, KU Leuven, Belgium. His research interests include UAV networks, mega-constellation networks, and integrated space-terrestrial networks.

\textbf{Hai Zhu}
is a PhD student at the Department of Cognitive Robotics at the Delft University of Technology. He received the B.E. (2015) and M.Sc. (2017) degrees from NUDT. His current research focuses on motion planning for multi-robot systems with an emphasis on real-time algorithms for aerial vehicles and planning under uncertainty. He received the IEEE ICRA best paper award on multi-robot systems in 2019.

\textbf{Lei Yang}
received his Ph.D. degree from NUDT, China, in 2008. He is currently a professor at NUDT. His current research interests are focused on space communication networks, onboard computer, spacecraft system modeling and simulation.

\textbf{Xiaoqian Chen}
received his Ph.D. degree from NUDT, China, in 2001. He is a professor at NUDT and National Innovation Institute of Defense Technology. His current research interests include spacecraft systems engineering, advanced digital design methods of space systems, and multidisciplinary design optimization.

\textbf{Sofie Pollin} 
is an associate professor with the Electrical Engineering Department, KU Leuven, Belgium. Her research centers around networked systems, which require networks that are ever more dense, heterogeneous, battery powered, and spectrum constrained.

\textbf{Evgenii Vinogradov}
is with the Department of Electrical Engineering, KU Leuven, Belgium, where he works on wireless communications with unmanned aerial vehicles (UAVs) and UAV detection. His research interests include UAV communications and multidimensional radio propagation channel modeling.

\end{document}